\documentclass[10pt, conference, a4paper]{IEEEtran}
\ifCLASSINFOpdf
\usepackage[pdftex]{graphicx}
\else
\fi
\usepackage{bytefield}
\usepackage[bookmarks=false]{hyperref}

\usepackage{units}

\PassOptionsToPackage{bookmarks=false}{hyperref}

\begin{document}
%
%
\title{Enabling Soft Vertical Handover for MIPv6 in OMNeT++}

\author{
\IEEEauthorblockN{Atheer Al-Rubaye, Ariel Aguirre, Jochen Seitz}
\IEEEauthorblockA{Communication Networks Group\\
Technische Universit\"at Ilmenau, Germany\\
\{atheer.al-rubaye, ariel.aguirre, jochen.seitz\}@tu-ilmenau.de}
}

\maketitle

\begin{abstract}

Switching connectivity over multiple wireless interfaces of a mobile node is essential when performing handover between heterogeneous networks. In such a communication scenario, session continuity as experienced by the user can be enhanced if soft handover is enabled through the concept of make-before-break. However, some of the available networking protocols need to be modified to support this feature. This work gives an insight into the supplementary modules we implemented and the modification conducted in MIPv6 model of OMNeT++ to facilitate soft handover and data offloading for mobile nodes.  
  
\end{abstract}


\section{Introduction}
\begin{figure} [ht]
\centering
\includegraphics[width=0.48\textwidth]{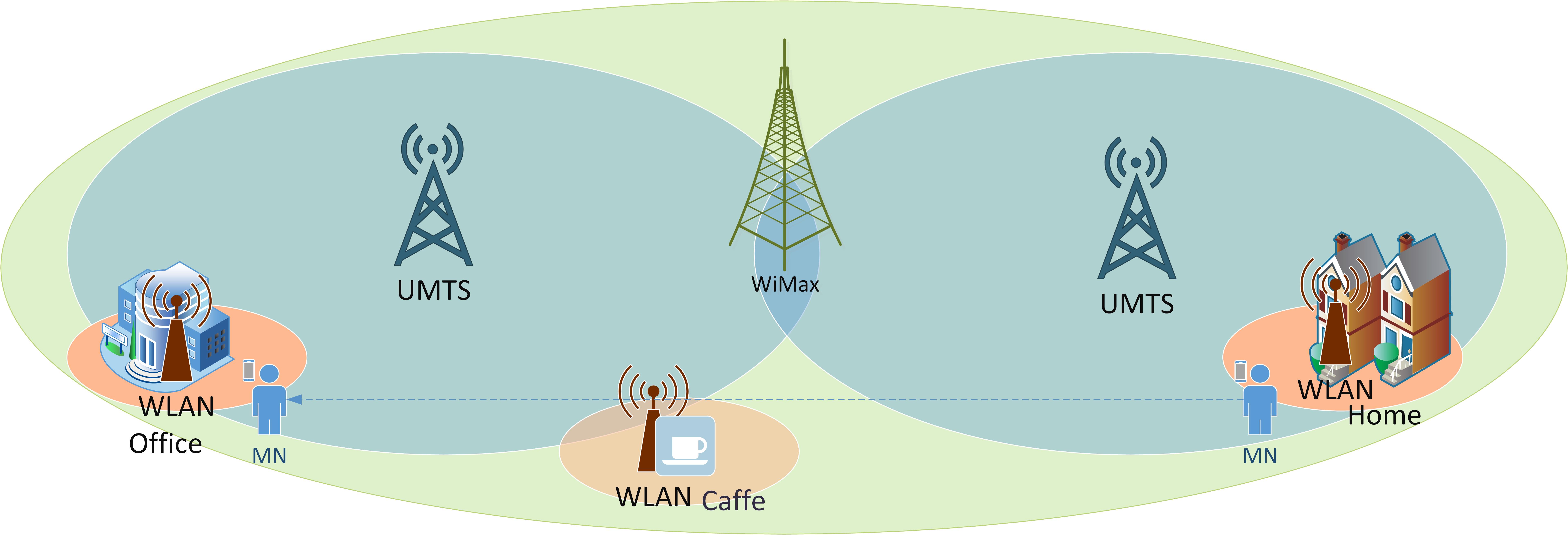}
\caption{Vertical Handover}
\label{fig:vho}
\end{figure}

Nowadays, heterogeneous communication networks coexist and overlap over areas so frequently. Considering the increasing availability of powerful phones equipped with multiple interfaces for different technologies, it can be employed to switch connectivity in between through a process known as vertical handover (VHO). A simple VHO scenario is start moving from home while having an active stream running, passing through the city, and reaching office, as illustrated in figure \ref{fig:vho}.
However, to offer a more sophisticated handover (HO), session continuity is a major factor to consider. A widely implemented protocol to handle the identification of mobile nodes upon roaming between networks of different subnet address (layer 3 HO) is mobile IP in its two versions 4 and 6 \cite{MIPv6_rfc3775, MIPv4_rfc3344}.

In this work, we give an insight into the modules and procedures we developed to add support of multiple interfaces utilization and provide session continuity to the solutions that are based on mobile IP in OMNeT++. Decision making and address resolution phases of handover are out of the scope here, where the main focus is on the management of interfaces connectivity. The concept of make-before-break is therefore implemented to provide a soft switching of traffic between the available wireless interfaces/networks.
An extension to our previous work in \cite{dinat15} is presented to support mobile IP version 6 (MIPv6) and adopt its functionality in xMIPv6 model \cite{zarrar} of INET v2.6 framework in OMNeT++ \cite{omnet++}. The support of the availability of multiple wireless interfaces with possible connectivity switching in between motivates further investigations and, hence, modifications to implemented protocols and modules. \cite{dinat15} implements a link layer controller module (vhoCtrler) within a compound model named IPCoManager to support soft handover in IPv4-based MNs. IPCoManager includes also handover decision algorithms in a more sophisticated form and an agent module that manages address resolution, as explained in \cite{dinat15}, these are however out of focus here. 

To enable this functionality for MIPv6 modules in INET, a set of modifications and adaptations are necessary to be conducted in xMIPv6 and in the vhoCtrler as well. Some are related to overcoming some limitations in xMIPv6, to enable more realistic mobility scenarios, like to have many handovers between the home and foreign networks when moving back and forth in between, and some are related to how IP address assignment and update mechanism should be carried out with the existence of more than one wireless interface.  
Enhancements in handover in terms of QoS parameters are investigated through simulations that emphasize the difference between a hard and a soft handover, and their impact on running sessions. 

This work is still under development and related modules may need further enhancements and modifications therefore, the code will be available after a successful finish of the PhD research work of the correspondent author.

The rest of the paper is organized as follows: section II gives an overview of MIPv6, in section III, an insight into the implemented and modified modules to support soft switching between interfaces is given, section IV describes the simulation scenarios and discusses the measurements, finally, the work is concluded in section V.

\section{Overview of MIPv6}
The protocol MIPv6 enables an IPv6 mobile node to be identified and reachable at any time, regardless to its point of attachment to the Internet.
In MIPv6, an MN is always identified through a fixed address called Home Address (HoA), which is derived from the home network IPv6 prefix. Once it is visiting a foreign network, the MN is reachable through a Care-of-Address (CoA), which is formed based on the prefix announced in that network. At any time, the anchor point of the MN is an entity in the home network, which the standard names Home Agent (HA).
MIPv6 enables the MN and HA to keep a binding of the CoA and HoA. Each time a MN changes its point of attachment to the Internet, it notifies the HA to update the binding with a new CoA. In such a way, packets sent by a correspondent node (CN) to the MN, are sent to the home network, intercepted by the HA, which then forwards them to the MN's CoA. In the reverse, packets sent by the MN to the CN are routed directly to it if the firewall in between permit \cite{MIPv6_rfc3775}.


\section{Implementation of Management Concept}

\subsection{Management at the Link Layer}

The controller module in \cite{dinat15} is a controlling entity to all wireless interfaces at the link layer. It receives requests from the underlaying interfaces for permissions to associate to its relevant networks when available in range. 

Regardless of the interface type, its highest management entity must be modified to acquire a permission before associating to its corresponding network. For example, the Agent module inside the wireless Ieee80211 compound module of INET will need permission from the vhoCtrler before allowing underlaying modules to associate to the related access point. It may also receive a command to disconnect an association with an access point if a handover decision is made and a switching process has taken place in favor of another interface.

The controller may consult a decision module for a selection, which compares between the current connected interface/network and the requesting one. However, the decision criteria and algorithm implementation are out of the scope here.
If a request is permitted, vhoCtrler releases the old connection, but only when the new one is confirmed to be associated, configured and hence, ready to undertake traffic.

\begin{figure} [ht]
\centering
\includegraphics[width=0.25\textwidth]{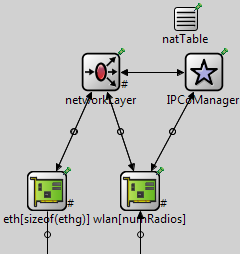}
\caption{Management of Interfaces}
\label{fig:crosslayer}
\end{figure}

To hold/retrieve the identities and parameters of multiple interfaces, the vhoCtrler module employs a structure for interface/network attributes. Three objects are defined accordingly, these namely are; serving, previous and candidate networks/interfaces for currently connected, previous and under examination interfaces respectively. The identities include the interface ID and the associated IP address. The parameters include attributes of the related access point like the received signal strength, the signal to noise ratio, and some other that might be transmitted by the network service provider like cost of service and limit of supported mobility speed. It includes also context information of an interface like the available credit (in data volume or monitory units) and hence, the expected bit rate of the possible multiple networks associations. The parameters are updated periodically with each received beacon/advertisement in relevance to each interface, and during the application run to enable a sophisticated decision making, which might be based on multiple attributes. 

It is important to notify here that the vhoCtrler does not deal with data traffic, but only exchanges messages with the aforementioned modules for purpose of management. Figure \ref{fig:crosslayer} illustrates the implemented module within the TCP/IP model of INET.

\subsection{Modifications at the Network Layer}

The need for modifications on the network layer in general, and the xMIPv6 in specific were necessary, because the availability of and the switching between multiple wireless interfaces was not tested before, and due to some missing functionality while employing a mobility model beyond the one implemented in the INET example of xMIPv6.

When a mobile node associates to a new wireless network, it starts receiving router advertisements (RA) from the router in that network.
The process on the received RA packets in the neighbor discovery procedure needed to be modified in order to make home-relevant information available to the serving interface, which is achieved by adding two pieces of information to the related serving wireless interface in the interface table. The first is the home network information and the second is the home address, but in tentative
mode. When the xMIPv6 class need to employ the addresses of the home agent and home address when a binding update (BU) is required to be sent, it queries the vhoCtrler to identify the interface that is labeled as the serving one momently, and retrieve the home information from the IPv6InterfaceData class accordingly.
 
Upon a handover, the routing table needed to be updated by deleting routing information related to the previous interface so, 
before the mobile node may employ its IPv6 address in the new network, the duplicated address detection (DAD) process is invoked. Once the scope of this address changes to global, a notification of this change is received by the vhoCtrler, which updates then the mobile node's routing table by removing the default route and prefix associated with the previous interface. This assures that the BU will be routed through the newly connected interface using a topologically correct address.
The MIPv6 registration process follows as described in standard \cite{MIPv6_rfc3775}.
For the case when an MN initiates a communication session while away from home, we have updated the original IPv6 procedures. Now, as in this case the MN and the home agent have built a bidirectional tunnel between them, we have filled the source IP address of the inner datagram with its home address, as described in the protocol standard in \cite{MIPv6_rfc3775}. This update is required in order to enable the corresponding node (CN) to return packets to the mobile node through the Home Network.



\section{Simulations and Measurements}

\begin{figure} [ht]
	\centering
	\includegraphics[width=0.45\textwidth]{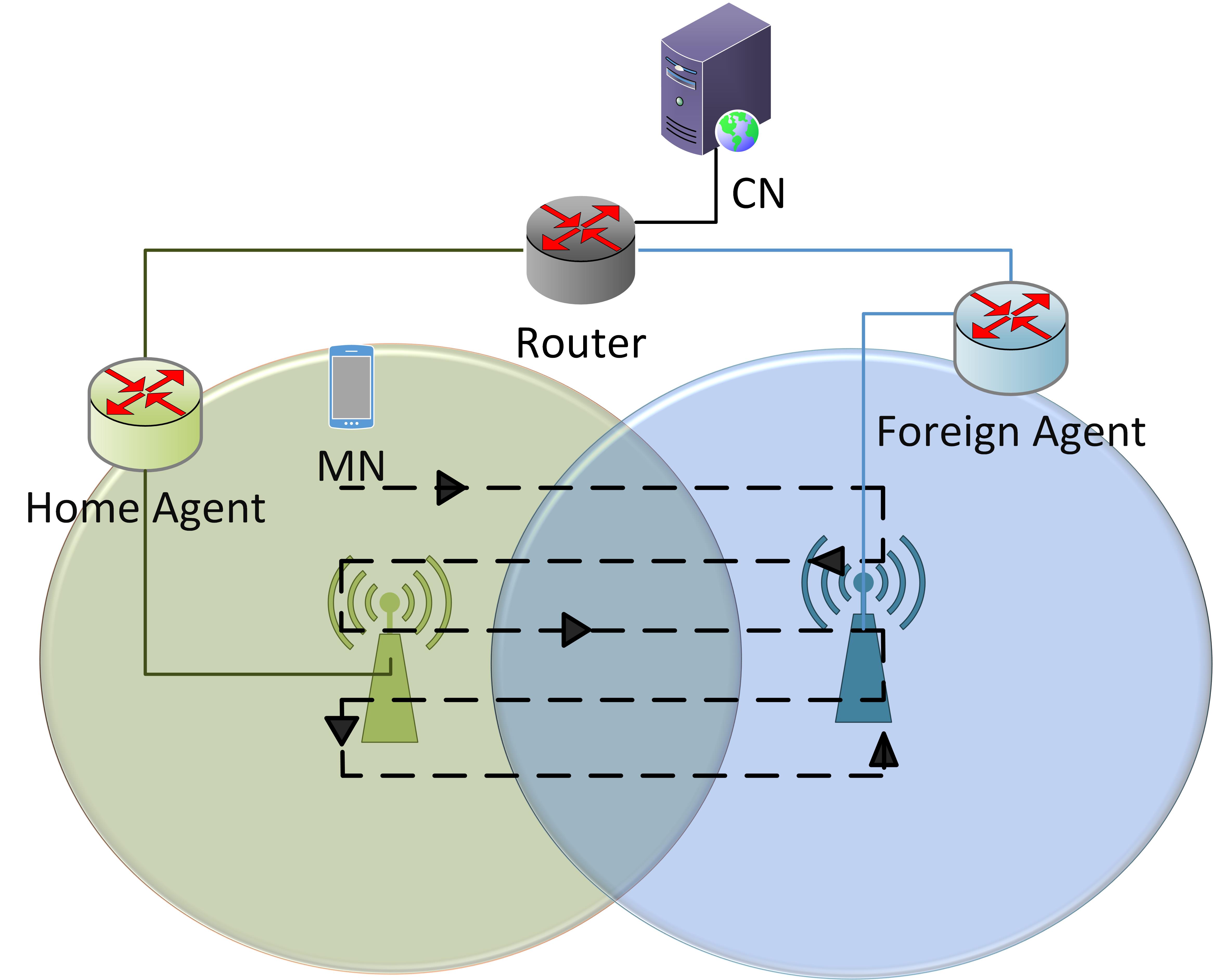} 
	\caption{Simulation Network Topology}
	\label{figs:toppology}
\end{figure}

In our simulated scenario, we consider a topology of two networks, one acts as the home network and the other as the foreign network in a free space environment. The access point of each is configured with the same transmission power but, on two different channels. The mobile node starts a video/VOIP traffic with the CN and moves in the meanwhile between the two networks to perform several handovers. As mobility model, we employ the TractorMobility provided in INET, which provides a path similar to a tractor moving through a field with specific number of rows, as illustrated in figure \ref{figs:toppology}. 

Two cases are considered, one with the MN using the basic version of MIPv6 of OMNeT++, which we refer to as Hard HO, and the second using our vhoCtrler module as well, which we refer to as Soft HO. The MN is equipped with a single wireless interface (Wi-Fi IEEE 802.11 with 2 Mbps bit rate) for the first case while, with two interfaces for the second case. In this case, each interface is configured to associate to only to one of the networks in the topology such that, the first interface connects only with the access point of the home network while, the second connects only with the access point of the foreign network. Roaming therefore requires switching of the traffic over the two interfaces in this scenario. 
Two real-time applications are employed; video stream and voice over IP. Sending rates of 0.5 and 2 Mbps were used for the video traffic, while a 5 ms play out delay and 20 ms packetization interval for the VOIP traffic.

The simulation was repeated five times with the same seed, using different mobility speed values in each (1, 2, 4, 8 and 10 mps). The simulation time differs for each run (2000 to 200 seconds) in proportion to the mobility speed in order to have the same number of handovers (10 times) in each run. The effect of the handover process is observed on the packet loss rate, which is a major parameter for real-time applications.

\begin{figure} [ht]
	\centering
	\includegraphics[width=0.5\textwidth]{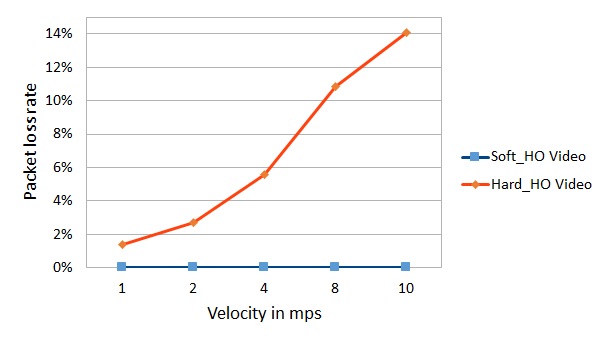} 
	\caption{Packet loss rate in the Video traffic case}
	\label{figs:video}
\end{figure}

\begin{figure} [ht]
	\centering
	\includegraphics[width=0.5\textwidth]{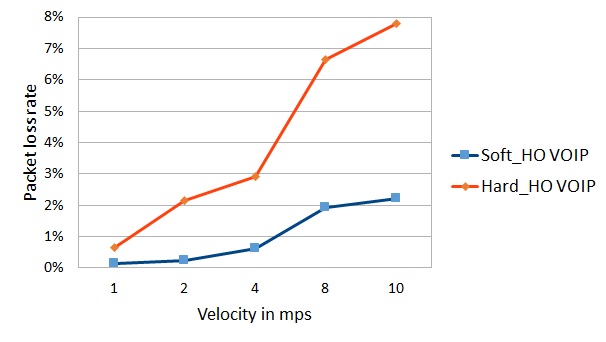} 
	\caption{Packet loss rate in the VOIP traffic case}
	\label{figs:voip}
\end{figure}

Figures \ref{figs:video} and \ref{figs:voip} illustrate the packet loss rate (for the video and VOIP traffic respectively). We notice a significant enhancement presented by soft handover when enabled for MIPv6, especially at higher velocities. However, for the VOIP application, the packets are not sent as a stream, but as talk spurs therefore, lately arrived packets might be also considered as lost in this case. 

\begin{figure} 
\centering
\includegraphics[width=0.5\textwidth]{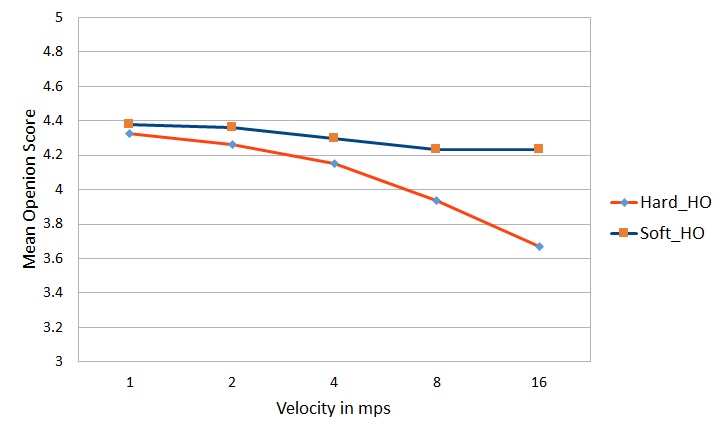}
\caption{MOS of the VOIP traffic case}
\label{fig:mos}
\end{figure}

The Mean Openion Score (MOS) measured by OMNeT++ for the VOIP application shows relatively similar performance at low speeds for both HO schemes but, better values can be noticed in higher speeds for the Soft HO as compared to the hard one, as can been seen in figure \ref{fig:mos}.

Handover in Soft HO case takes place while being in the coverage overlap area between the two networks and switching is executed only when the second connection is really created (make-before-connect). In Hard HO case, handover takes place only when the MN is on the edge of the connected network, where it starts associating its single wireless interface to the upcoming network after notifying the absence in the received beacons of the connected one.

\section{Conclusions and Future Work}

To enable soft handover in the MIPv6 model implemented in INET/OMNeT++, we needed to adopt the xMIPv6 model and our controlling entity as well, which we implemented in an earlier work of us for IPv4-based models. A set of modifications to the xMIPv6 were necessary to support the availability of multiple interfaces and to enable a correct operation of the protocol according to the standard. 
In the two tested applications, a soft switching between two interfaces in a coverage overlap area has highly reduced the number of lost packets upon a handover. 
However, an overall examination of the code in comparison to the xMIPv6 model is still necessary to ensure compatibility with the protocol standard and a correct functionality within more sophisticated simulation scenarios.
Adding features of extended versions of MIPv6, like HMIPv6 and PMIPv6 to xMIPv6 can be considered as a significant contribution to the OMNeT++ community as a future work. 

\bibliographystyle{IEEEtran} 

\bibliography{A.OMNeTCS16}

\end{document}